\documentclass[twocolumn,showpacs,preprintnumbers,amsmath,amssymb]{revtex4}

\usepackage{graphicx}
\usepackage{dcolumn}
\usepackage{bm}
\usepackage{subfigure}
\begin{document}

\title{Systematic {\em ab initio} study of the phase diagram of epitaxially strained SrTiO$_3$}
\author{Chien-Hung Lin, Chih-Meng Huang and G. Y. Guo\footnote{Electronic address:
gyguo@phys.ntu.edu.tw}}
\address{Department of Physics, National Taiwan University, Taipei, Taiwan 106, Republic of China}
\date{\today}

\begin{abstract}
We use density-functional theory with the local-density approximation to study 
the structural and ferroelectric properties of SrTiO$_3$ under misfit strains.  
Both the antiferrodistortive (AFD) and ferroelectric (FE) instabilities are 
considered by calculating all the phases predicted by 
Pertsev {\em et al.} [Phys. Rev. B {\bf 61}, R 825 (2000)] 
based on the phenomenological Landau theory. 
The rotation of the oxygen octahedra and the movement of 
the atoms are fully relaxed within the constraint of a fixed in-plane lattice constant. 
We find a rich misfit strain-induced phase transition sequence 
which is in overall agreement with the prediction by Pertsev {\em et al.}
and is obtained only when the AFD distortion is taken into account. 
Nevertheless, the calculated locations of the phase boundaries are different
from the prediction by Pertsev {\em et al.}
We also find that compressive misfit strains induce ferroelectricity in the
tetragonal low temperature phase only whilst tensile strains induce ferroelectricity
in the orthorhombic phases only. The calculated FE polarization
for both the tetragonal and orthorhombic phases increases monotonically
with the magnitude of the strains.
The AFD rotation angle of the oxygen octahedra
in the tetragonal phase increases dramatically 
as the misfit strain goes from the tensile to compressive strain region whilst it decreases
slightly in the orthorhombic (FO4) phase. 
This reveals why the polarization in the epitaxially strained SrTiO$_3$
would be larger when the tensile strain is applied, since the AFD distortion is found
to reduce the FE instability and even to completely suppress it in the small strain region.
Finally, our analysis of the average polar distortion and the
charge density distribution suggests that both the Ti-O and Sr-O layers contribute 
significantly to the FE polarization.
\end{abstract}

\pacs{77.22.Ej, 77.84.Dy, 77.80.-e}
\maketitle

\section{Introdution}
The strontium titanate (SrTiO$_3$, STO) is a prototypical perovskite compound.
A mechanically free bulk STO will go through two soft mode-driven phase 
transitions as the temperature is lowered.
First, there is the cubic-to-tetragonal antiferrodistortive (AFD) phase transition 
at 105 K\cite{105,rot}, which results from the phonon induced instability 
of the $R$ point of the 
Brillouin zone (BZ) boundary and involves the rotation of the TiO$_6$ octahedra.
This is then followed by the ferroelectric (FE) phase transition driven by the instability 
of the zone-center soft phonon mode. However, the FE transition is not 
observed in pure STO due to quantum fluctuations\cite{qflu}.
Therefore, it is regarded as an incipient ferroelectric.

In the perovskite ferroelectrics, it is well known both experimentally and theoretically 
that the polarizations are strongly coupled to strain\cite{uwe}, and thus the properties 
such as FE transition temperature and polarization magnitude are quite sensitive 
to external stress.
Recently, using techniques such as molecular beam epitaxy and pulsed-laser deposition, 
it has become possible to grow high-quality perovskite films, and has attracted a great 
deal of attention.
In particular, Haeni {\em et al. }\cite{nature} recently reported a giant shift 
in FE transition temperature 
to 293 K for a STO film under biaxial strains. They grew STO films on a 
(110)-oriented DyScO$_3$ substrate, corresponding to a biaxial tensile strain 
of 0.8 \%. The results indicate that even 
under small strains, a nonpolar STO film can be dramatically transformed to a polar state.

Several theoretical studies have provided insight into the structural 
properties and temperature 
dependence of phase transitions in an epitaxial STO film.\cite{pertsev,srtio3_1,srtio3_2,die05}
A phenomenological study of STO thin films under strain by 
Pertsev {\em et al.}\cite{pertsev}, 
based on a Landau model fitted to experimental data from bulk phases, predict a rich 
temperature-strain phase diagram.
The effects of in-plane strain and an applied electric field 
on the dielectric properties of 
STO epitaxial thin films were calculated by Antons {\em et al.}\cite{srtio3_1} using 
density-functional theory with the local-density approximation (LDA). 
They find that the polarization can be tuned by the in-plane lattice parameter.
Recent {\em ab initio} calculations by Hashimoto {\em et al.}\cite{srtio3_2} confirmed
that the ferroelectricity could be induced by applying biaxial strains and also indicated 
a mechanical origin of the ferroelectricity.

However, in all previous first-principles calculations on the expitaxial STO films,
the AFD instability is neglected. 
This is thought of as an approximation to the high temperature phases.
Consequently, many epitaxial strain induced low temperature phases were unable to
be investigated in the previous calculations.
In this paper, we investigate the effects of the misfit strains on 
the structure and ferroelectricity in the epitaxially strained STO 
with both the FE and AFD instabilities taken into account.
We use parameter-free first-principles LDA calculations to systematically 
analyze all the predicted phases by Pertsev {\em et al.} \cite{pertsev}
As a result, we find a rich epitaxial strain induced low temperature phase transition
sequence that was unobtainable in the previous {\em ab initio} calculations. 
We further show the effects of the strains on the AFD instability itself, and hence 
the difference in the orientation of the oxygen octahedra and also the polarization 
magnitude compared with the case when the AFD instability is ignored.

The rest of this paper is organised as follows.
In Sec. II, we describe the computational technique used and summarize the 
salient results of the previous phenomenological analysis. 
In Sec. III, we present and discuss the results of our calculations 
for the epitaxial STO films under misfit strains. 
We begin with the determination of the theoretical stable structures of all 
the possible phases, and then proceed to relate the FE and 
AFD instabilities to the epitaxial strain and also the underlying
electronic band structure. Finally, we give a summary in Sec. IV.

\section{Theory and computational method}
Our {\it ab initio} calculations for the STO were performed using accurate 
full-potential projector augmented wave (PAW)\cite{paw1,paw2} method, as implemented in the 
VASP\cite{vasp1,vasp2} package. They are based on density-functional theory with the 
LDA. PAW potentials are used to describe the electron-ion interaction, with 
the 4$s$, 4$p$, 4$d$ and 5$s$ states of Sr, 3$d$ and 4$s$ states of Ti, and
2$s$, 2$p$ of O being treated as valence states.
A conjugate-gradient minimization scheme is used to 
minimize the Kohn-Sham energy functional, with a plane-wave cutoff energy 
of 400 eV used throughout.  All calculations are performed with 
a 6$\times$6$\times$6 Monkhorst-Pack $k$-point mesh. To establish 
minimum-energy configurations the total forces acting on the atoms are converged to less 
than 0.01 eV/\AA. Our calculations are carried out at the theoretical equilibrium lattice 
constant of 3.87 \AA\ which is $\sim$0.9 \% less than the experimental value 3.905 \AA, 
under different strains between $2 \%$ and $-2 \%$.

\subsection{Structure optimization}
We begin by systematically performing optimizations of the five-atom unit 
cell of the high-temperature phases (FT1, HT and FO1) and the twenty-atom 
supercell of all the other possible phases involving the rotations of the oxygen octahedra 
proposed by Pertsev {\em et al.}\cite{pertsev} 
In Fig. 1, 
we illustrate the supercell which is chosen to describe the AFD 
instability and give a schematic representation of the atomic displacements 
for the $R$ point [$(111)\pi/a$] phonon mode.
The epitaxial strain on the structural properties of STO is modeled 
by relaxing either the five-atom or twenty-atom cell of bulk STO 
under the constrained in-plane lattice constants, 
differing from the theoretical cubic lattice constant by fractions ranging 
from -2 \% to 2 \%. A description of these phases is given in Table I. 
We have worked out the space groups of all the phases which are also listed
in Table I.
For each phase, starting from a structure in which the symmetry is established by 
appropriately displacing the Ti and O atoms, we relax the atomic positions until 
the atomic forces fall below a small threshold of 0.01 eV/\AA.


Na Sai and Vanderbilt\cite{AFD} have considered the possible rotations of the oxygen 
octahedra in the STO but only at zero strain. Here we consider the possible 
rotations of the oxygen octahedra
in the STO under the misfit strains. 
We focus on the ground state tetragonal phase obtained by freezing in an 
AFD phonon mode at the $R$ point of the Brillouin zone (BZ) boundary.
This phonon mode corresponds to the rotation of the TiO$_{6}$ octahedra in 
the opposite directions from one unit cell to the next. 
For the order parameter $q_{1}$, we take, e.g.,  the $x$-axis as the rotation axis, 
and adopt a tetragonal supercell with the lattice vectors of 
length $\sqrt{2}a$, $\sqrt{2}a$, and $c$ along 
the $[011]$, $[0\overline{1}1]$, and $[200]$ directions, respectively.
Similarly, cyclic permutations of the lattice vectors would give rise to the other 
two rotation axes.

\subsection{Thermodynamic theory analysis}
Pertsev {\it et al.}\cite{pertsev} used a phenomenological Landau theory to describe the 
mechanical substrate effect on equilibrium states and phase transitions
in epitaxial STO thin films, and developed a misfit strain-temperature phase
diagram. Instead of dealing with thin films directly, they computed the structure of the 
bulk material strained to match a given cubic 
substrate with square surface symmetry. On the basis of a Landau theory fitted 
to experimental data from bulk phases, the thermodynamic 
description may be developed from the power-series expansion of the Helmholtz 
free-energy density $F$ in terms of temperature $T$, polarization $\bf{P}$ and 
order-parameter $\bf{q}$, which corresponds to linear oxygen displacements 
correlated with the rotation of the oxygen octahedra around one of fourfold 
symmetry axes, and lattice strain $S$, defined as $S=(a-a_0)/a$ (where $a$ is 
the effective lattice constant of the substrate and $a_0$ denotes the lattice 
constant of free-stranding cubic cell).
By minimizing the free energy $F(T,S,\bf{P},\bf{q})$ with respect to the 
components of the polarization and structural order parameter, they predicted 
the equilibrium thermodynamic states of STO films under various temperature-strain 
conditions. The misfit strain-temperature phase diagram of STO is more complex than 
BaTiO$_3$ (BTO) because the coexistence of two coupled AFD and FE instabilities 
in this crystal. 
The results are listed in Table \ref{pertsev}.
Apart from the high temperature tetragonal (HT) phase, they contain purely 
structural tetragonal and orthorhombic states (ST and SO), purely 
ferroelectric tetragonal and orthorhombic phases (FT1 
and FO1), and four coupled states (FT2, FO2, FO3 and FO4).

In this paper, using parameter-free total-energy methods based on density 
functional theory, we map out the equilibrium structures of the STO as a 
function of epitaxial strain at zero temperature, and find out the effect 
of misfit strain on the magnitude and orientation of the polarization.
The electric polarization is calculated using the Berry phase method\cite{berry1,berry2}.

 
\section{Results and discussion}
Because of the competition between ferroelectric and structural instabilities, 
the phases in the STO are more complex than the BTO case\cite{batio3}.
For clearly displaying the calculated energy for each phase as a function 
of the misfit strain, we separate them into two classes, which are separated 
by a large energy difference, as shown in Figs. 2 and 5. 
One contains the high temperature phases, which do not involve the rotation 
of the oxygen octahedra (FT1, HT, and FO1).
The other includes the low temperature phases with the coupling between the 
polarization and the rotation of the oxygen octahedra.

\subsection{High temperature phases}




The total energy for each of the three high energy phases as a 
function of misfit strain is shown in Fig. 2. 
For large compressive strains, the stable structure is the FT1 phase; 
for large tensile strains, the FO1 phase is prefered. 
At the misfit strain of around -0.1 \%, there is a continuous transition 
from the polar FT1 phase to the nonpolar HT phase. 
As the in-plane misfit strain is increased to 0.3 \%, 
the STO transits from the HT phase to the polar FO1 phase.
In Fig. 3, 
we show the calculated polarizations for the 
FT1 and FO1 phases versus the misfit strain.
The polarization increases dramatically with the magnitude of the misfit strain. 
For compressive strains larger than -0.1 \%, the FT1 phase 
has the polarization along $[001]$ whilst the FO1 has zero polarization. 
As the misfit strain gets close to zero, a so-called paraelectric gap 
within which all the phases exhibit zero polarization, appears (Fig. 3).  
For tensile strains larger than 0.3 \%, the FO1 phase acquires a large polarization 
along $[110]$, whilst the polarization of the FT1 phase becomes zero.

In Fig. 4, 
we show the calculated atomic displacements 
as a function of misfit strain. 
Under large compressive strains, the atoms in the FT1 phase relax 
only along the [001] direction. As the in-plane strain decreases, 
a second-order FT1 $\rightarrow$ HT phase transition 
follows, and the magnitude of the atomic 
displacements along [001] gradually diminishes.
While the displacements along [001] in the FT1 phase vanish around the zero strain,
the displacements in the [110] direction in the FO1 begin to grow as we enter 
the tensile strain region.
With the increasing tensile strain, the displacements in the $xy$ plane
in the FO1 phase continue to grow monotonically.
Near zero misfit strain, the atomic displacements change from along 
the $z$-direction to along an $xy$ plane direction, 
and the change of the orientation of the polarization follows.
From these results, we can see that the in-plane misfit strain make 
the polarization to be a monotonically increasing function of the 
magnitude of the strain, though the polarization direction switches 
in the paraelectric gap.  

The above results are similar to that of the {\em ab initio} LDA 
calculations in Ref. \onlinecite{srtio3_1}.  Nevertheless, there are 
some quantitative differences. For example, our estimation of the 
paraelectic gap (between -0.1 \% and +0.3 \% ) is narrower than that 
previously obtained (between -0.75 \% and +0.54 \% in Ref. \onlinecite{srtio3_1}), 
and this may be attributed to the different potentials used 
in the present and previous calculations. Furthermore, because 
the rotations of the oxygen octahedra were not included in the analysis
in Ref. \onlinecite{srtio3_1}, the results of Ref. \onlinecite{srtio3_1} correspond 
to our results for the high temperature FO1, HT and FT1 phases only.

\subsection{Low temperature phases}
Due to the interaction between the FE polarization and the AFD structural instability, 
the low temperature phases are more complex than the high temperature ones. 
In Fig. 5, 
we show all the low energy phases predicted in 
Ref. ~\onlinecite{pertsev} for the misfit strains between -2 \% and 2 \%.
Roughly speaking, we get nearly the same features as the high temperature case: 
the presence of two 
wide misfit strain ranges in which the STO becomes a ferroelectric. 
This is due to the coupling between the polarization and the misfit strain. 
In the films grown on the compressive substrates, a FE 
phase with polarization along the out-of-plane orientation ([001]) appears, 
while in the case of the tensile 
substrates, the polarization direction becomes in-plane ([110]).
Nevertheless, there are some subtle differences between the high
and low temperature cases.
First, the structures in the low temperature cases are more stable 
than in the high temperature ones, because they have a significantly
lower total energy (Figs. 2 and 5). 
Under large compressive strains, the FT2 phase 
has the lowest energy among the low temperature phases. 
When the misfit strain is decreased, the STO transits from the polar FT2 
phase to the nonpolar ST phase and the 
two energy curves overlap in a rather wide range of tensile 
strains (Fig. 5). 
In contrast, for the high temperature phases,
the energy curve of the nonpolar HT phase overlaps with the 
polar FT1 phase in the tensile strains only, but with the polar 
FO1 phase in both the tensile and compressive strains.
For large tensile strains, the FO4 phase is energetically most favorable.
There are some other stable phases occuring only in a 
small strain region near zero.
The energy curves of the FO2 and SO are almost the same, 
and their trends are also similar to the FO4 phase in the small 
strain region, but split off from the FO4 with increasing tensile strain (Fig. 5). 
Similarly, the curve of the FO3 phase lies on the ST phase one in the compressive strain 
but splits off from it when grown on the tensile substrate.
As the compressive or tensile strain increases, the
structural symmetry is broken, thereby splitting the energy degeneracy.
However, in the small strain region, the total energies for many different
phases are almost in overlap with each other. Nevertheless, accurate 
distinction between different stable phases can be achieved by calculating 
the second derivative of the total energy with respect to 
the strain. The strain divisions obtained this way, between FT2 and ST, ST and FO3, 
FO3 and FO2, FO2 and FO4 are -0.558 \%, -0.221 \%, 0.197 \%, and 0.442 \%, respectively.
Remarkably, this rich low temperature phase-versus-strain sequence is  
generally consistent with the prediction of the phenomenological theory 
by Pertsev {\em et al.}\cite{pertsev} The main 
difference is the precise location of the low temperature phase boundaries.
In particular, we predict that the FO3 phase is located in the strain region
between -0.22 \% and 0.20 \% whilst it falls in the strain region 
from around -0.023 \% to -0.005 \% according to Pertsev {\em et al.}\cite{pertsev}
The FO2 is predicted to stride the zero strain in Ref. \onlinecite{pertsev},
instead of the FO3 phase. 
Since the FO2 and FO3 phases have different polarization directions and
also different order parameter {\bf q}, this difference can perhaps be easily
examined by further experiments.
Of course, our calculations are for the zero and very low temperatures only. 
Consequently, we cannot discuss the temperature effect on 
the epitaxial STO structures which may lift the energy degeneracy, and hence
the misfit strain-temperature phase diagram.



In Fig. 6, 
we show the calculated polarization (including both ionic and 
electronic contributions) of the FT2, FO2, FO3 and FO4 phases as a function of misfit strain.
Interestingly, the energy curves of all the low temperature phases
converge to the same energy at the zero strain (Fig. 5). 
However, away from this narrow window near the zero strain, the lowest energy phase is
the FT2 under the large compressive strains and the FO4 in the tensile strain region
(Fig. 5). 
Therefore, we concentrate on these two phases only below.
The results are similar to the high-temperature phase one.
The polarization is along [001] and [110] for the compressive and tensile strains, respectively.
It exhibits a monotonic increase with the magnitude of the misfit strain.
There is also a paraelectric gap centered at the zero strain, now between -0.4 \% and 0.4 \%.
This gap is larger than the high temperature one. Nevertheless, 
the size of this gap should be the upper bound because we have not considered the 
other FE phases in the paraelectric region.
The polarization for the FO2 and FO3 phases becomes zero when the misfit strain drops
below 0.2 \% and 0.3 \%, respectively. Therefore, the overall paraelectric gap for the 
low temperature phases is between -0.4 \% and 0.2 \%.
Interestingly, a comparison of the polarization-versus-strain curves for the FT1 phase
in Fig. 3 
and the FT2 phase in Fig. 6 
reveals 
that the AFD instability reduces the polarization in the large compressive strain
region, and completely suppresses it in the small strain region, thus enlarging the
paraelectric gap. Note that the atomic structures of the FT1 and FT2 phases are almost the same
except that the FT2 phase can accommodate the AFD rotations of the oxygen
octahedra whilst the FT1 phase cannot. Indeed, the AFD distortion as characterized
by the rotation angle of the oxygen octahedra in the FT2 is strong in the compressive
strain region, as shown in Fig. 8 below.


We now further quantify the strength of the FE and AFD distortions.
In the FE case, let us define the average polar distortion along 
the $i$-direction, $p_i$, in the 1$\times$1 unit cell of each atomic layer.
For the TiO$_2$ layers,
\[p_i=\Delta_i(Ti)-\frac{\Delta_i(O_1)+\Delta_i(O_2)}{2},\]
and for the SrO layers,
\[p_i=\frac{\Delta_i(Sr_1)+\Delta_i(Sr_2)}{2}-\Delta_i(O),\]
where $\Delta_i$ are the atomic displacements in the $i$-direction relative to 
the undistorted symmetric 1$\times$1 perovskite structure.
The computed results describing the strength of the FE distortion in 
each atomic layer are ploted in Fig. 7. 
Note that for the 
ST and SO phases, the average polar distortion is zero for all
the strains, and therefore, is not shown in Fig. 7. 

It is clear from Fig. 7 
that in the compressive strain region, 
the average polar distortion in the 
FT2 phase is larger in the TiO$_2$ layers than in the SrO layers.
Interestingly, in the FO4 phase, it is the opposite, i.e., the average distortion
in the tensile strain region is larger in the SrO layer than in the TiO$_2$ layer.
Around the zero strain, the average distortion is significant only in the
TiO$_2$ layer in the FO3 and FO2 phases. This indicates that in the vicinity 
of the zero strain, only the FO2 and FO3 phases have a significant polarization,
thereby giving a microscopic explanation to the results of the
previous thermodynamical analysis\cite{pertsev} (Table I).
Note also that the average polar distortion along the in-plane direction in
the tensile strain region in the FO3 and FO4 phases is larger than along the 
z-direction in the compressive strain region in the FT2 phase.
This suggests that the ionic contribution to the polarization in the FO4 phase
in the tensile strain region would be larger than in the FT2 phase in
the compressive strain region, and explains the calculated polarizations
displayed in Fig. 6. 


The strength of the AFD distortion is characterized by the TiO$_4$ 
rotation angle $\theta$, in each TiO$_2$ layer. 
In Fig. 8, 
we present the calculated rotation angle of the oxygen octahedra as a 
function of misfit strain for the low-temperature phases due to the AFD instability.
The calculated tilting angle of the oxygen octahedra at zero strain (6$^\circ$) is 
overestimated in our calculations, as also in Ref. \onlinecite{AFD}, compared to 
the experimental value of 2$^\circ$\cite{angle}. This discrepancy between the 
{\it ab initio} calculations and the experiments may be caused by the LDA error 
or by the quantum fluctuations~\cite{zho96} which should 
reduce the rotations of the oxygen octahedra, or by both.
The results nonetheless provide us some qualitative insight into how 
oxygen octahedra rotate under the applied strains.
Fig. 5 
shows that under compressive strains, the FT2, ST and FO3 are the 
low energy phases and the rotation angle of the TiO$_6$ octahedra along [001] 
increases with the magnitude of the misfit strain. 
In contrast, under the tensile strains, the FO4, FO2 and SO become the low energy phases 
(Fig. 5) 
and the rotation axis changes to lie in the $xy$ plane. 
In the FO2 and SO phases, the rotation angle of the TiO$_6$ octahedra 
again increases with the magnitude of misfit strain (Fig. 8). 
Interestingly, the rotation of the oxygen octahedra in the FO4 phase 
is nearly independent of applied strain (Fig. 8). 
Note that there are two equivalent rotation axes along [100] and [010], respectively,
for the FO4 phase. Therefore the linear displacements of the oxygen atoms due to the rotations
and the rotation angles are equal for the two axes. In Fig. 8, 
we thus 
show only the rotation angle for either [100] or [010].
Interestingly, Figs. 6 and 7 
show that
the FE instability as manifested in the electric polarization
and the average polar distortion, in both the FT2 phase in the large compressive 
strain region and the FO4 phase in the large tensile strain one,
grow as the magnitude of the strains is increased,
though it is stronger in the FO4 phase than in the FT2 phase. 
In contrast, the behavior of the AFD instability as manifested in the rotation of the
oxygen octahedra, is different. Fig. 8 
shows that the tilting 
angle of the oxygen octohedra in the FT2 (also ST and FO3) phase increases dramatically as 
the misfit strain goes from the strong tensile region through zero strain to
the strong compressive region, while it remains nearly as a relatively 
small constant for the FO4 phase. This explains why the calculated polarization is 
larger for the FO4 phase under the tensile strain 
than the FT2 phase when the AFD instability which tends to
suppress  the FE instability, is taken into account.

\subsection{Charge density and polarization}

In order to better understand the chemical bonding and also the origin of the
electric polarization in the strain induced FE phases, we examine the calculated
charge density distributions.
Figure 9 
shows the contour plots of the charge density 
differences between calculated charge densities and those obtained 
by a superposition of the free atomic charge densities for 
the FT2 phase under compressive strain of -1.6 \% and the FO4 phase at tensile strain of 1.6 \%.
It can be seen from the calculated charge density difference distribution
on the (001) SrO planes in Fig. 9 
that in the both structures,
the charge is depleted in the inner region of the Sr sites, whilst, in contrast, 
there is considerable charge buildup on the O sites and also in the outer region
of the Sr sites. This indicates that the Sr-O bonding is predominantly ionic.
The ionic Sr-O bonding can be further seen from the fact that
the charge distrubtion at the Sr sites are nearly spherical. 
This is because the Sr 5{\em s} valence electrons have moved to the
neighboring O atoms which have high electron affinitiy, 
with the spherical symmetry ionic cores left behind.
On the other hand, the bonding between the Ti and O atoms are strongly
covalent, as can be seen from the calculated charge density difference distributions
on the (001) and (100) TiO planes in Fig. 9. 
There is pronounced
charge buildup in the vinicity of the Ti-O bond center whilst there is significant
charge depletion on both the Ti and O sites. 

Figure 9 
also shows that the charge density distribution on both 
the (001) SrO and TiO planes in the FT2 phase has a fourfold symmetry at the 
O and Ti sites, respectively. As a result, there is no electric polarization 
in the $x$-$y$ plane. In contrast, on the (001) TiO plane in the FO4 phase
[Fig. 9(d)] 
the charge buildup is much stronger 
in the central region of the two right Ti-O bonds than in the
left two ones, and it leans towards the vertical down direction
in all the four Ti-O bond regions, thus resulting
in an electric polarization along the [110] direction (vertical up direction).
Clearly, the charge density distribution on the (001) SrO plane
in the FO4 phase also lean downwards, suggesting that the
SrO layers also contribute to the polarization, being in conflict with
common wisdom.
On the (100) TiO planes [Fig. 9(e-f)],
the strong charge buildup
in the FT2 phase clearly lean downwards and is symmetric with respect to
the vertical line through the center O atom in the FT2 phase, 
whilst, in contrast, it leans
towards left and is symmetric with respect to the horizontal line through
the center O site in the FO4 phase. As a result, the FT2 phase has a
polarization along [001] whilst the FO4 phase exhibit a polarization
along [110].
 
\section{Summary}
In this paper, we have performed {\em ab initio} LDA calculations  
for the STO under various misfit strains.
A unique part of this work is that we have considered both the high-temperature and 
low-temperature phases of the STO proposed by Pertsev {\em et al.}\cite{pertsev} such that 
both the AFD and FE structural instabilities have been investigated,
unlike previous {\it ab initio} studies in which  
the AFD rotations of the oxygen octahedra have been neglected.
As a result, we find a rich misfit strain-induced low temperature phase transition sequence
which is in overall agreement with the prediction by Pertsev {\em et al.}
Nevertheless, the calculated locations of the phase boundaries are rather different
from the prediction by Pertsev {\em et al.} 
In particular, we predict that the FO3 phase is located in the strain region
between -0.22 \% and 0.20 \% whilst the FO2 would stride the zero strain
according to Pertsev {\em et al.}
Since the FO2 and FO3 phases have different polarization directions and
also different order parameter {\bf q}, this difference can perhaps be easily
examined by the future experiments.
By analyzing the FE atomic displacements and the AFD rotations of the oxygen octahedra, 
we provide a comprehensive microscopic insight into the effects of both the FE and AFD 
instabilities on the ferroelectric properties of STO under the epitaxial strains. 
We also find that at low temperatures, compressive strains would induce FE polarization in the
FT2 phase only whilst tensile strains induce polarization
in the orthorhombic phases only. The calculated ferroelectric polarization
for both the FE tetragonal and orthorhombic phases increases monotonically
with the magnitude of the strains, with the polarization in the tensile strain region
being larger than in the compressive region. The AFD rotation angle of the oxygen octahedra
in the FT2 phase increases dramatically
as the misfit strain goes from the tensile to compressive strain region whilst it remains
nearly unchanged in the FO4 phase.
This reveals why the polarization in the epitaxially strained STO
would be larger when the tensile strain is applied, since we find that the AFD distortion
would reduce the FE instability and even completely suppress it in the small strain region.
Finally, by analysing the so-called average polar FE distortion and the
charge density distribution, we find that both the Ti-O and Sr-O layers contribute
significantly to the FE polarization.

\begin{acknowledgments}

The authors gratefully acknowledge financial supports from National Science Council
of the Republic of China. They also thank National Center for High-performance
Computing of the Republic of China for providing CPU time.\\
\end{acknowledgments}



\begin{table*}
\caption{\label{pertsev} The epitaxial SrTiO$_{3}$ phases predicted
by Pertsev {\em et al.} \cite{pertsev}
Nonzero components of the polarization $\bf{P}$ and the order parameter $\bf{q}$ are the
characteristics of the different phases in the epitaxial SrTiO$_3$ films grown on the
cubic substrates. Phases are labeled by the lattice symmetry (T = tetragonal, O = orthorhombic)
and also the type of the instability (S = structural, F = ferroelectric, H = high temperature).
Also listed are the space groups ($G_s$) for all the phases found in the present work. }
\begin{ruledtabular}
\begin{tabular}{cccccccccc}
Phase &HT&ST&SO&FT1&FT2&FO1&FO2&FO3&FO4 \\ \hline
$\bf{P}$& & & & $P_3$ & $P_3$ & $|P_1|=|P_2|$ & $P_1$ (or $P_2$) & $|P_1|=|P_2|$ & $|P_1|=|P_2|$ \\
$\bf{q}$ & & $q_{3}$ & $q_1$ (or $q_2$) & & $q_3$ & & $q_2$ (or $q_1$) & $q_3$ & $|q_1|=|q_2|$ \\
$G_s$ & P4/mmm & I4/mcm & Fmmm & P4mm & I4cm & Amm2 & Fmm2 & Ima2 & Ima2  
\end{tabular}
\end{ruledtabular}
\end{table*}

\newpage
\noindent {\bf Figure captions}\\

\noindent Fig. 1 (Color online) A [001] projected section of the SrTiO$_3$ lattice
illustrating the linear displacements of the oxygen ions associated with an
antiferrodistortive phonon mode at the $R$ point [$(111)\pi/a$] of
the Brillouin zone boundary (upper panel). The thick square frame indicates the supercell used
in our calculations. The lower panel is a schematic representation of the rotations
of the oxygen octahedra in the soft phonon-mode.\\

\noindent Fig. 2 (Color online) Calculated total energy of the high temperature
epitaxial SrTiO$_{3}$ phases versus misfit strain (see Table \ref{pertsev} for the
symbols of the phases).
The vertical dotted lines denote the boundaries between the paraelectric
and ferroelectric phases (see Fig. 3). \\

\noindent Fig. 3 Calculated polarization of the FT1 and FO1 phases versus misfit strain.
The polarizations for the FT1 and FO1 phases are along [001] and [110], respectively.
The vertical dotted lines denote the boundaries between the paraelectric
and ferroelectric phases.\\

\noindent Fig. 4 (Color online) Calculated atomic displacements $\Delta$ in the FT1 phase
(compressive strains only) and the FO1 phase (tensile strains only).
The subscripts label the directions of the displacements.\\

\noindent Fig. 5 (Color online) Calculated total energy of the low temperature
epitaxial SrTiO$_{3}$ phases versus misfit strain. The vertical dotted lines denote
the boundaries between the paraelectric and ferroelectric phases (see Fig. 6).\\

\noindent Fig. 6 (Color online) Calculated polarization of the FT2, FO2, FO3 and FO4 phases 
versus different strain. The polarization for the FT2 phase is along [001]; the polarization 
for the FO3 and FO4 phases is along [110]; the polarization for the FO2 phase is along [100]. 
The vertical dotted lines denote the boundaries between the paraelectric
and ferroelectric phases.\\

\noindent Fig. 7 (Color online) The average layer
ferroeletric polar distortion (see text), {\em p$_i$}, versus misfit strain,
in percentage of the lattice parameter.
Solid and dashed curves denote the distortions in the TiO$_2$ and
SrO layers, respectively.  The subscripts label the directions of the distortions.
Note that for the ST and SO phases, the average polar distortion is zero for all
the strains.\\

\noindent Fig. 8 (Color online) Calculated rotation angle of the oxygen octahedra
for all the low temperature epitaxial SrTiO$_{3}$ phases versus misfit strain.\\

\noindent Fig. 9 (Color online) Contour plots of the charge density differences
between the calculated charge densities and those obtained by a superposition of the
free atomic charge densities in different projected planes for the FT2 phase under
compressive strain of -1.6 \% and the FO4 phase at tensile strain of 1.6 \%.
(a) and (b) are for the (001) SrO$_2$ planes. (c) and (d) are for the
(001) TiO planes. (e) and (f) are for the (100) TiO planes. The dark (blue) curves
denote the positive contour levels with contour step of 0.025 e/\AA$^3 $and
the light (red) dashed curves denote the negative ones with contour step of 0.084 e/\AA$^3$.

\newpage

\begin{figure}
\includegraphics[width=6cm]{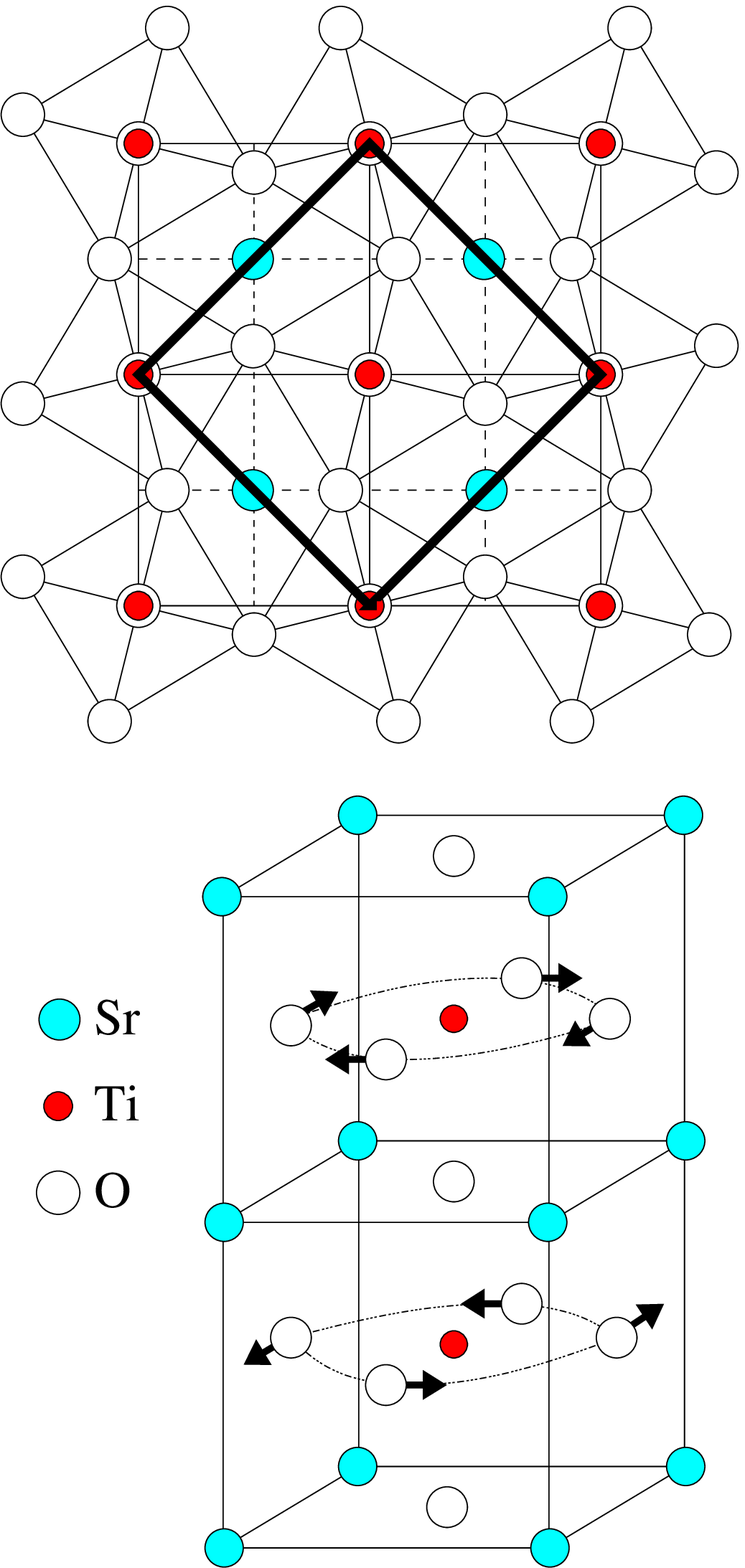}
\caption{\label{cell} }
\end{figure}

\begin{figure}
\includegraphics[width=8cm]{CHLinFig2.eps}
\caption{\label{e_high} }
\end{figure}

\begin{figure}
\includegraphics[width=8cm]{CHLinFig3.eps}
\caption{\label{p_high} }
\end{figure}
                                                                                                      
\begin{figure}
\includegraphics[width=6cm]{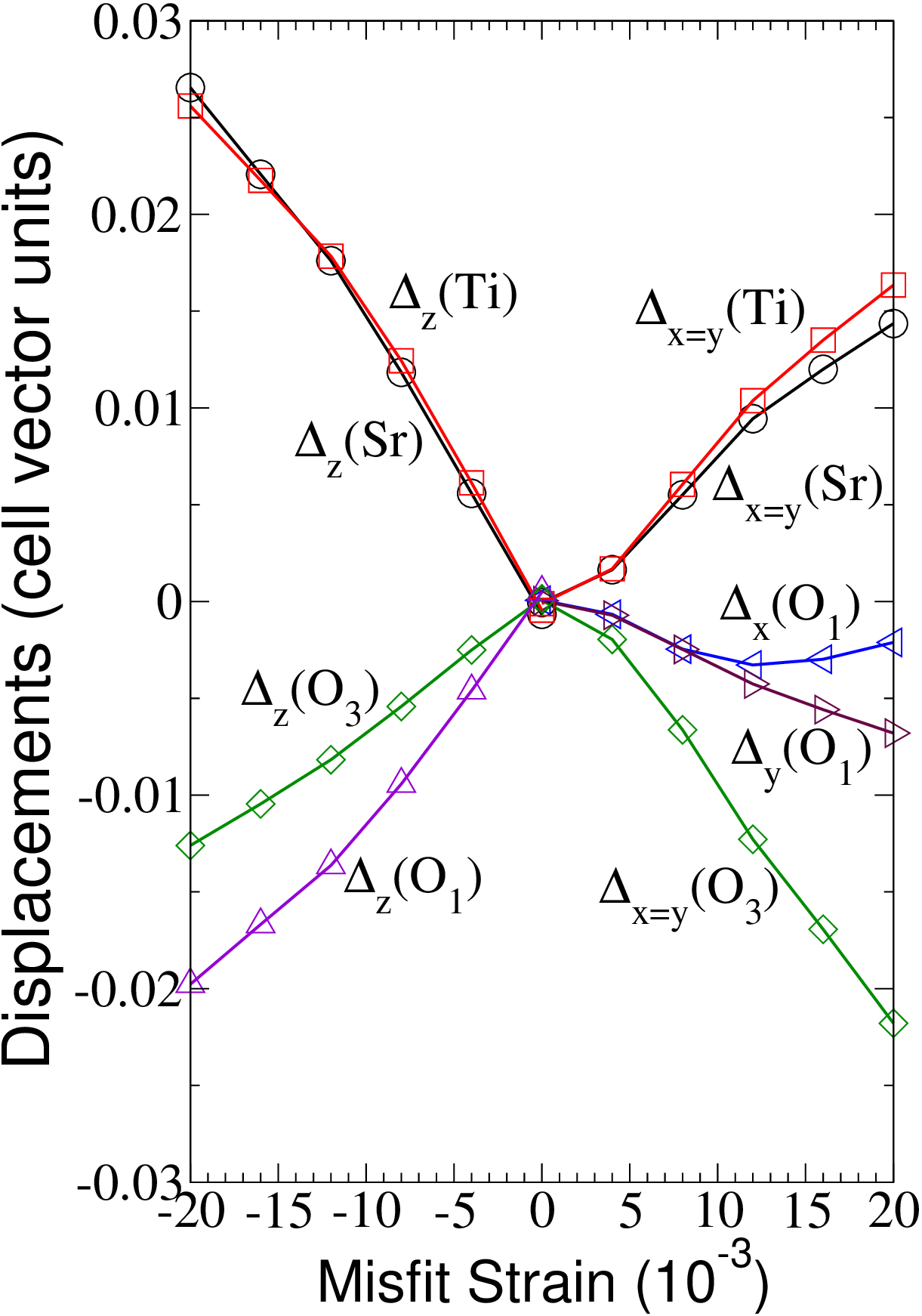}
\caption{\label{posi} }
\end{figure}

\begin{figure}
\includegraphics[width=8cm]{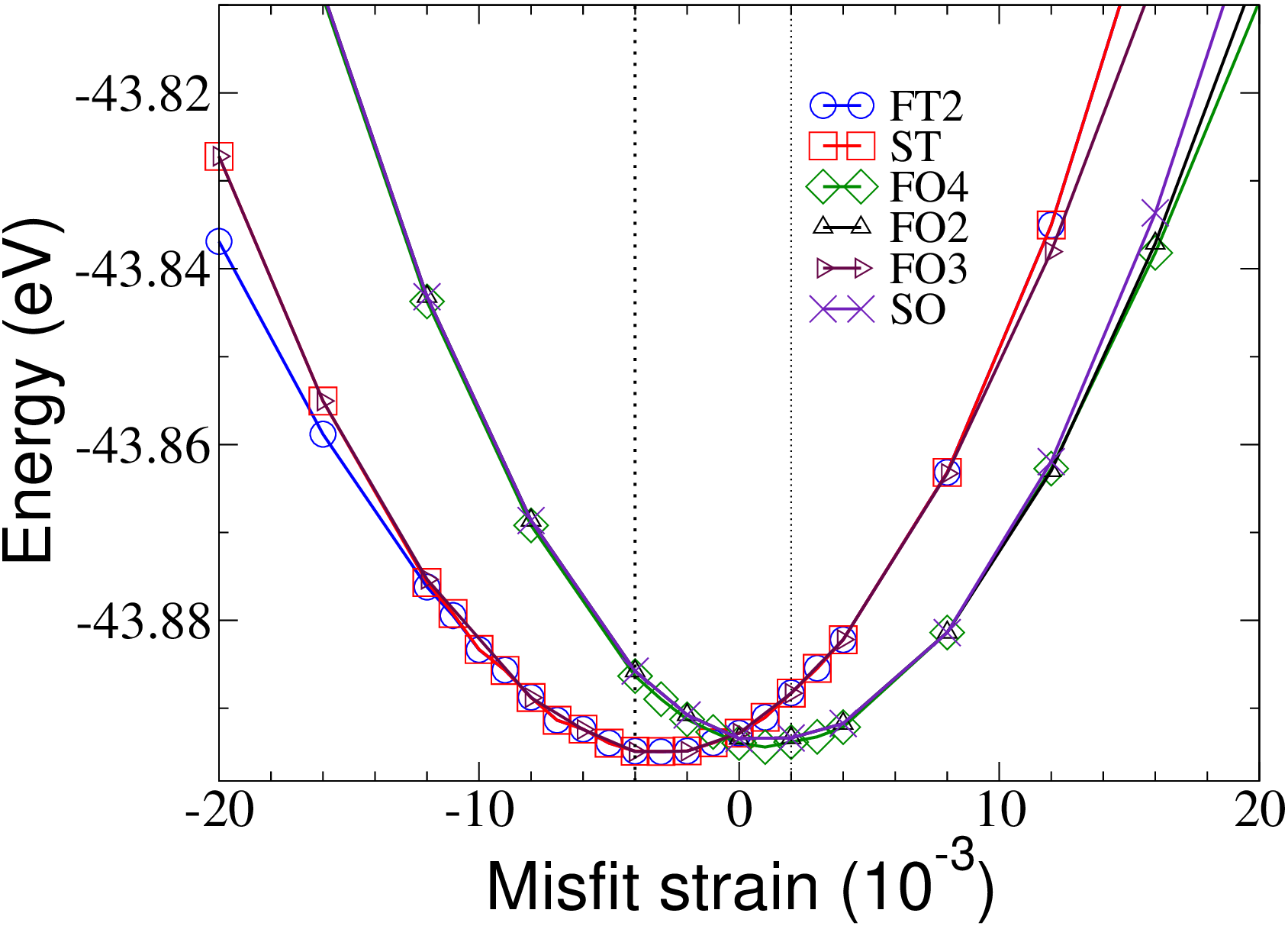}
\caption{\label{e_low} }
\end{figure}
                                                                                                        
\begin{figure}
\includegraphics[width=8cm]{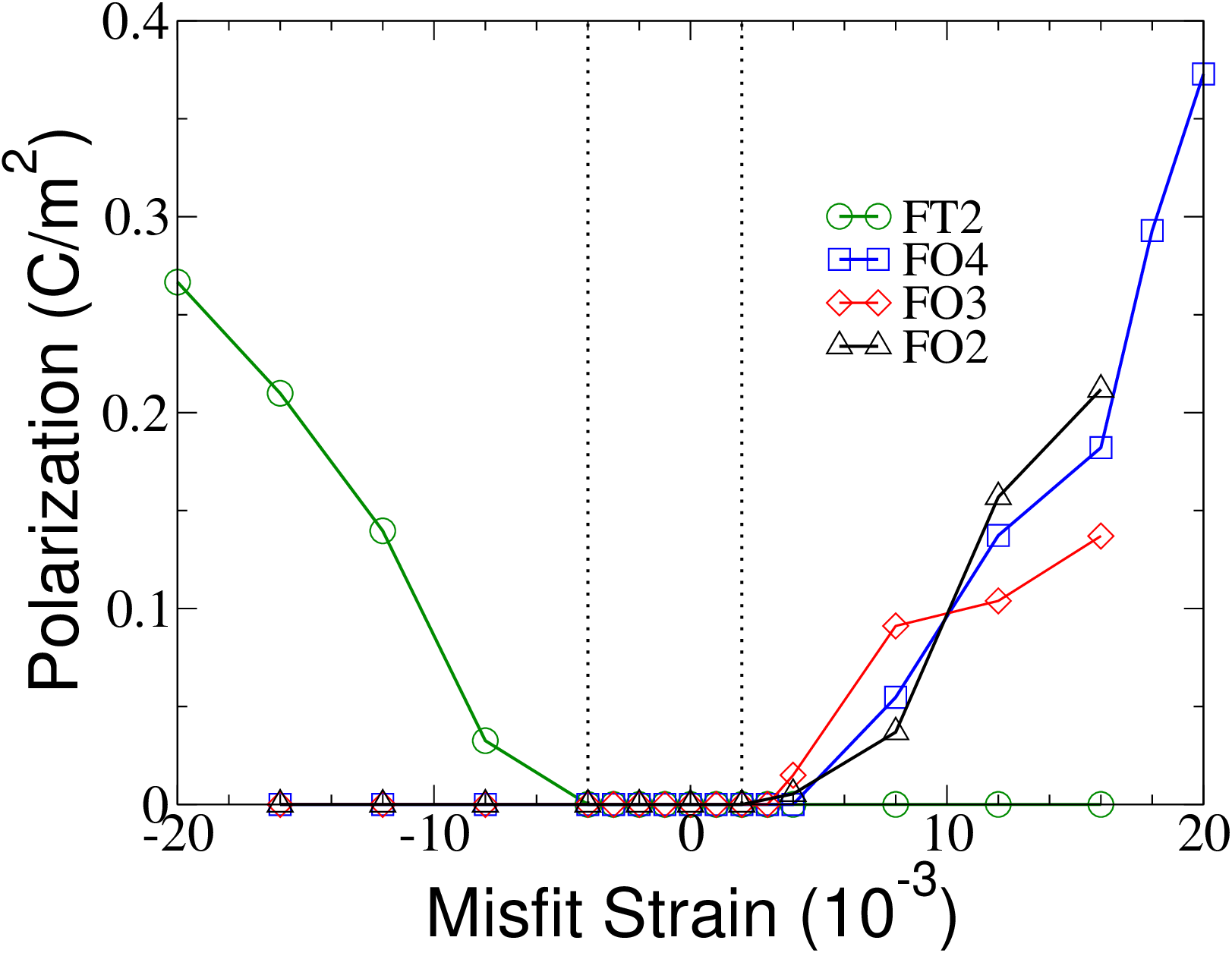}
\caption{\label{p_low} }
\end{figure} 
 
\begin{figure}
\includegraphics[width=8cm]{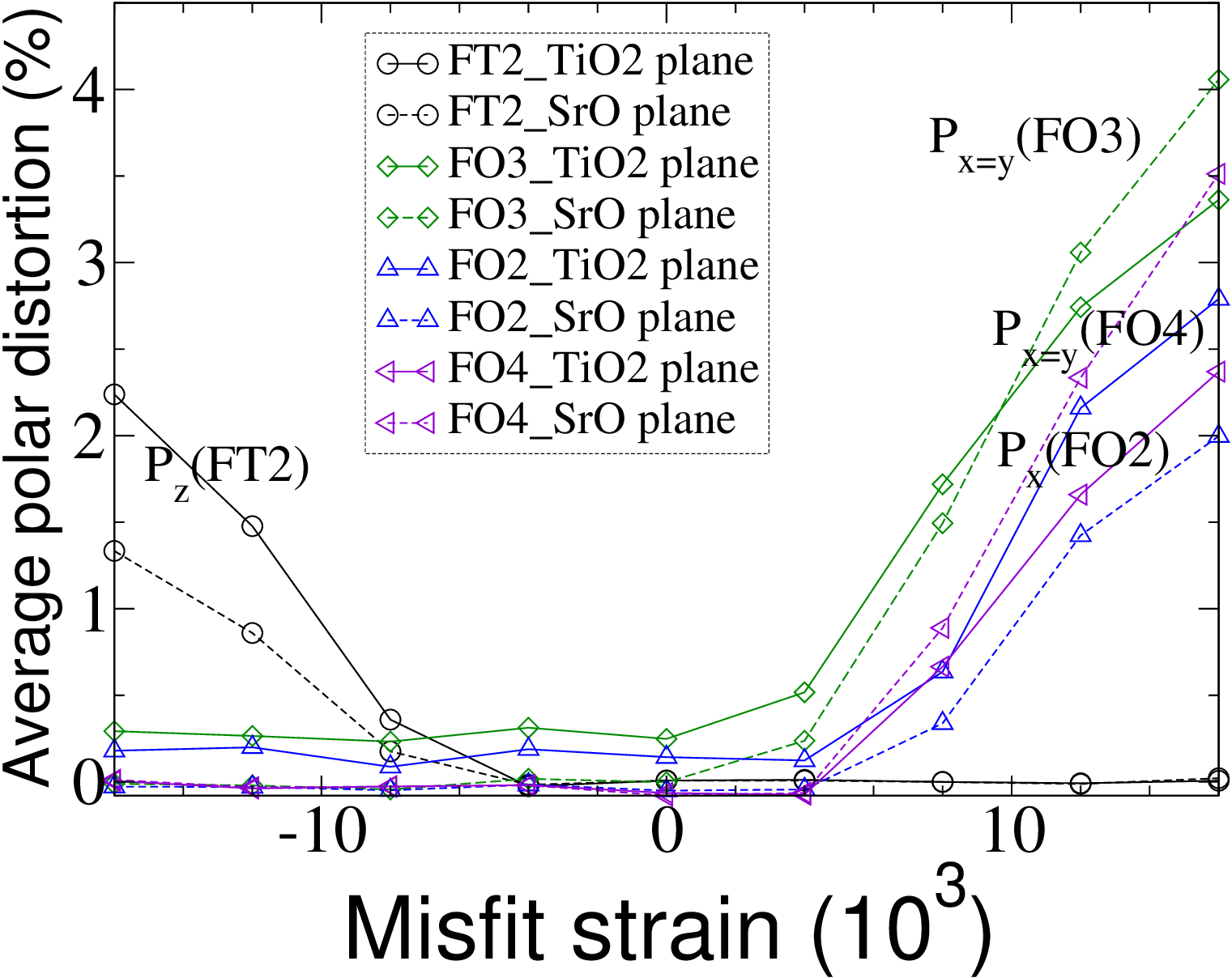}
\caption{\label{distortion} }
\end{figure}

\begin{figure}
\includegraphics[width=6cm]{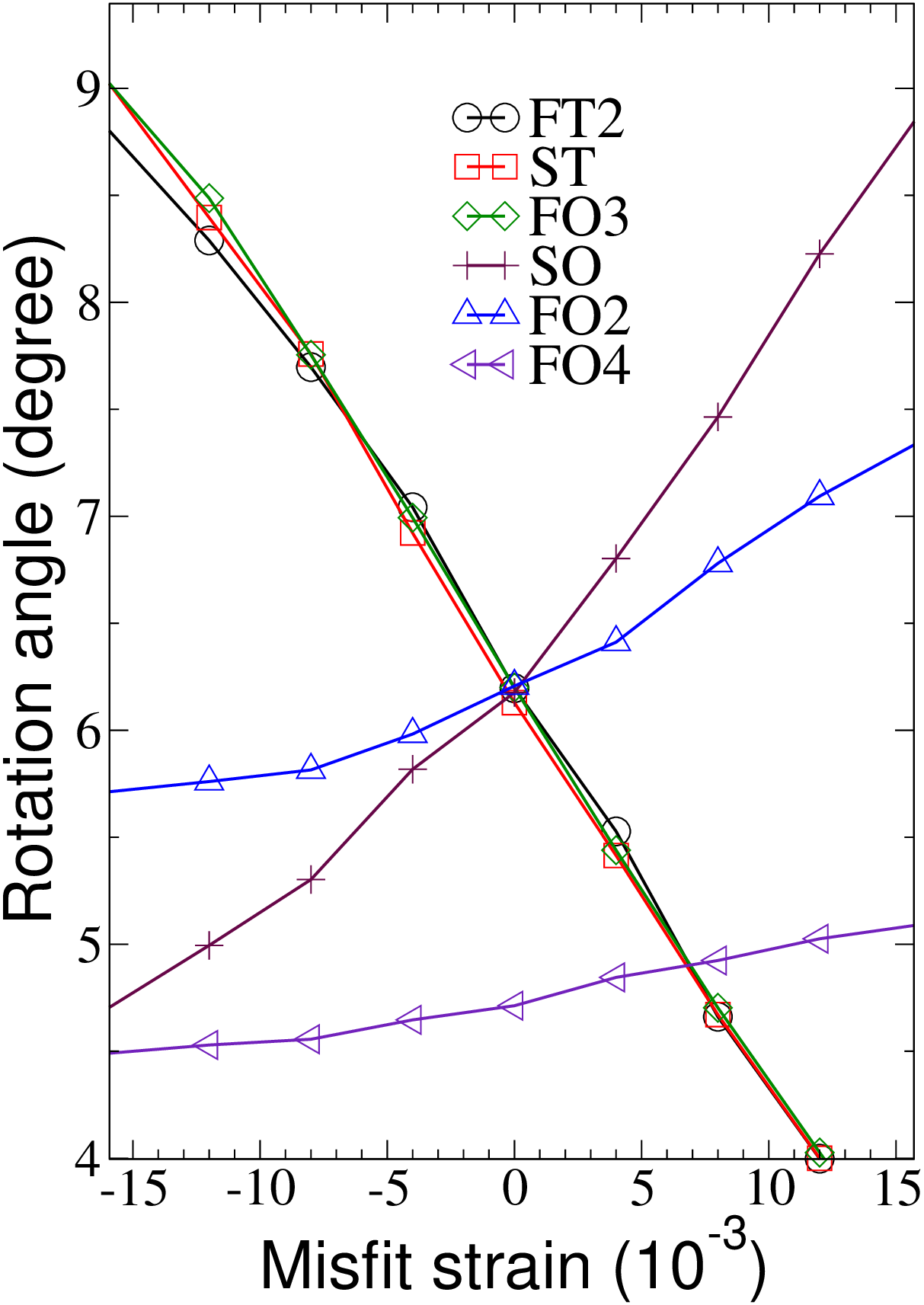}
\caption{\label{rot} } 
\end{figure}

\begin{figure}
\caption{\label{charge} }
\end{figure}

\end{document}